\documentclass[12pt,preprint]{aastex}
\usepackage{natbib}
\begin{document}

\title{Resolving the Sub-AU-Scale Gas and Dust Distribution in FU Orionis Sources}

\author{J.A. Eisner\altaffilmark{1,2} \&  L. A. Hillenbrand\altaffilmark{3}}

\altaffiltext{1}{Steward Observatory, University of Arizona, Tucson, AZ 85721}
\altaffiltext{2}{Alfred P. Sloan Research Fellow}
\altaffiltext{3}{Astrophysics Department, California Institute of Technology, Pasadena, CA 91125}


\keywords{stars:pre-main sequence---stars:circumstellar 
matter---stars:individual(FU Ori, V1057 Cyg, V1515Cyg)---techniques:spectroscopic---techniques:interferometric}

\begin{abstract}
We present Keck Interferometer observations of the three prototypical FU
Orionis stars, FU Ori, V1057 Cyg, and V1515 Cyg.   With a
spatial resolution of a few milli-arcseconds and a spectral 
resolution of $\lambda/\Delta \lambda= 2000$, our near-infrared
observations spatially resolve gas and
dust emission extending from stellocentric radii of $\sim 0.05$ AU to
several AU.  We fit these data with accretion
disk models where each stellocentric radius of the disk is represented
by a supergiant-type stellar emission spectrum at the disk temperature.
A disk model is consistent with the data for FU Ori, although we
require some local asymmetry in the disk.
For V1057 Cyg the disk model does not fit our data well, especially
compared to the fit quality achieved for FU Ori.  We
speculate that a disk wind may be contributing substantially to the
observed near-IR emission in this source.  The data for V1515 Cyg are
noisier than the data obtained for the other two objects, and do not
strongly constrain the validity of an accretion disk model.
\end{abstract}

\section{Introduction \label{sec:intro}}
The class of FU Orionis objects was initially defined by sudden,
large increases in optical-wavelength fluxes in 
FU Ori, V1057 Cyg, and V1515 Cyg
\citep{HERBIG66,HERBIG77}.   After these brightening events the
optical spectra of these objects resemble those of low surface gravity
(i.e., supergiant) stars.  However for V1057 Cyg, pre-outburst
spectra show that the progenitor resembled a higher surface gravity,
typical T Tauri type star \citep[e.g.,][]{HERBIG58}.  While
there are now many more members of the FU Ori class
\citep[e.g.,][]{RA10,MILLER+11}, the three prototypes remain very well-studied.

FU Ori outbursts are believed to result from periods of enhanced
accretion in circumstellar disks around young stars
\citep[e.g.,][]{HK85,HK96,ZHU+07}.  Outbursts may be triggered by thermal
instabilities \citep{BL94}, gravitational instabilities \citep{VB05}, a
combination of the two \citep{ALP01,ZHU+09}, or by interactions of a
disk with a planet or nearby stellar companion
\citep{LC04,CLARKE+05,RA04,PFALZNER08}.    Spectroscopic observations
of FU Ori stars have also been explained in the context of rapidly rotating stars
near the edge of stability \citep[e.g.,][]{HPD03}.  However
spectroscopic data at a range of wavelengths exhibit various features
that are more consistent with disks than stellar photospheres,
including highly broadened lines and poor cross-correlation
significance with single-temperature, fast-rotating giant star spectra
\citep[e.g.,][Hillenbrand et al. in prep.]{GAR08}.

The disk model is strongly supported by interferometric observations
that resolve the circumstellar emission at near-IR and mid-IR
wavelengths on scales $>100$ R$_{\odot}$.
Near- and mid-IR
observations of FU Ori itself can be explained in the context of
a geometrically thin accretion disk extending to stellocentric radii as
small as 5.5 R$_{\odot}$, inclined $55^{\circ}$ with respect
to the plane of the sky \citep{MALBET+05,QUANZ+06}.  The high inferred
disk temperature implies an accretion rate $>10^{-5}$ M$_{\odot}$
yr$^{-1}$, substantially higher than for typical T Tauri stars.
The disk model for FU Ori supported by the interferometric data is
also compatible with spectrophotometric data across a broad wavelength
range \citep[e.g.,][]{ZHU+08}.

However the disk model may not be the whole story, and other
components like extended envelopes or outflows may be important in
explaining the data for other FU Ori sources.
The sizes of near-IR emitting regions in
V1057 Cyg and V1515 Cyg are larger than expected from pure disk
models, suggesting the presence of scattered light from extended
structures \citep{MILLAN-GABET+06b}.  Indeed, these objects also
display excess long-wavelength emission that can not be reproduced
with geometrically thin disks \citep{KH91,GREEN+06,ZHU+08}.  To
explain these data, models with both disks and circumstellar
envelopes have been constructed.

A key prediction of accretion disk models is differential rotation.  A
Keplerian disk rotates faster at smaller stellocentric radii, and
shorter-wavelength spectral lines are expected to have broader
linewidths.  Indeed, such behavior has been observed in a number of FU
Ori objects \citep[e.g.,][]{HHC04,GAR08,ZHU+09}.  While often
difficult to reconcile with available interferometric data, alternative
explanations for these line profiles have been proposed
\citep[e.g.,][]{HPD03}.  Spatially resolved near-IR spectra can
determine the location of the gas directly, and provide a direct test
of the rotating disk model.

Recent enhancements to the Keck Interferometer (KI)
now enable spatially resolved spectroscopy of the near-IR emission
from circumstellar disks \citep{EISNER+10,POTT+10}.  
Here we present KI observations of the
three prototypical FU Ori objects: FU Ori,
V1057 Cyg, and V1515 Cyg.  Our data probe the entire $K$-band and are
sensitive to emission from the Br$\gamma$ transition of hydrogen
and the ro-vibrational overtone transitions of CO.  We determine the
relative spatial distributions of line and continuum emission in these
systems, providing critical tests of circumstellar disk models.

\section{Observations and Data Reduction \label{sec:obs}}
KI is a fringe-tracking long baseline near-IR Michelson
interferometer combining light from the two 10-m Keck apertures 
\citep{CW03,COLAVITA+03}.  Each of the 10-m apertures is equipped with
a natural guide star adaptive optics (NGS-AO) system that corrects
phase errors caused by atmospheric turbulence across each telescope
pupil, and thereby maintains spatial coherence of the light from the
source across each aperture.  Optical beam-trains transport the light
from each Keck aperture down into a tunnel connecting the two Kecks
and to a set of beam combination optics.

We used the ``self phase referencing''(SPR)  mode of KI, implemented as
part of the ASTrometric and phase-Referenced Astronomy (ASTRA) program
(see Woillez et al. 2011 for a description of the SPR mode).  SPR allows
deep integrations in a spectroscopic channel, which enables higher
dispersion without loss of signal-to-noise.  We use a grism with
$\lambda/\Delta \lambda = 2000$ to disperse the light.  
Note, however, that spectra are fully Nyquist sampled only at a
resolution of $\sim 1000$.   Near-IR spectral lines in FU Ori sources
may be as broad as $\sim 100$ km s$^{-1}$
\citep[e.g.,][]{GAR08,BISCAYA+97}, and thus the spectral resolution of
KI is reasonably matched to the observed systems.

Neon lamp spectra and Fourier Transform Spectroscopy are
used to determine the wavelength scale for each night of observed data.
While the entire $K$-band falls on the detector, vignetting in the
camera leads to lower throughput toward the band  edges.  The
effective bandpass of our observations is approximately 2.15 to 2.36
$\mu$m.

We obtained Keck Interferometer (KI) observations of V1057 Cyg on
2009 July 15, of V1515 Cyg on 2010 July 21, and of FU Ori on 2010
November 25 (all dates UT).   On each night we also observed
spatially unresolved stars of known spectral types, used to calibrate
the instrumental response.  The calibrator stars are HD 192985 and HD
195050 for V1057 Cyg; HD 194206 and HIP 102667 for V1515 Cyg; and HD
37147 and HD 42807 for FU Ori.

\begin{figure}
\plotone{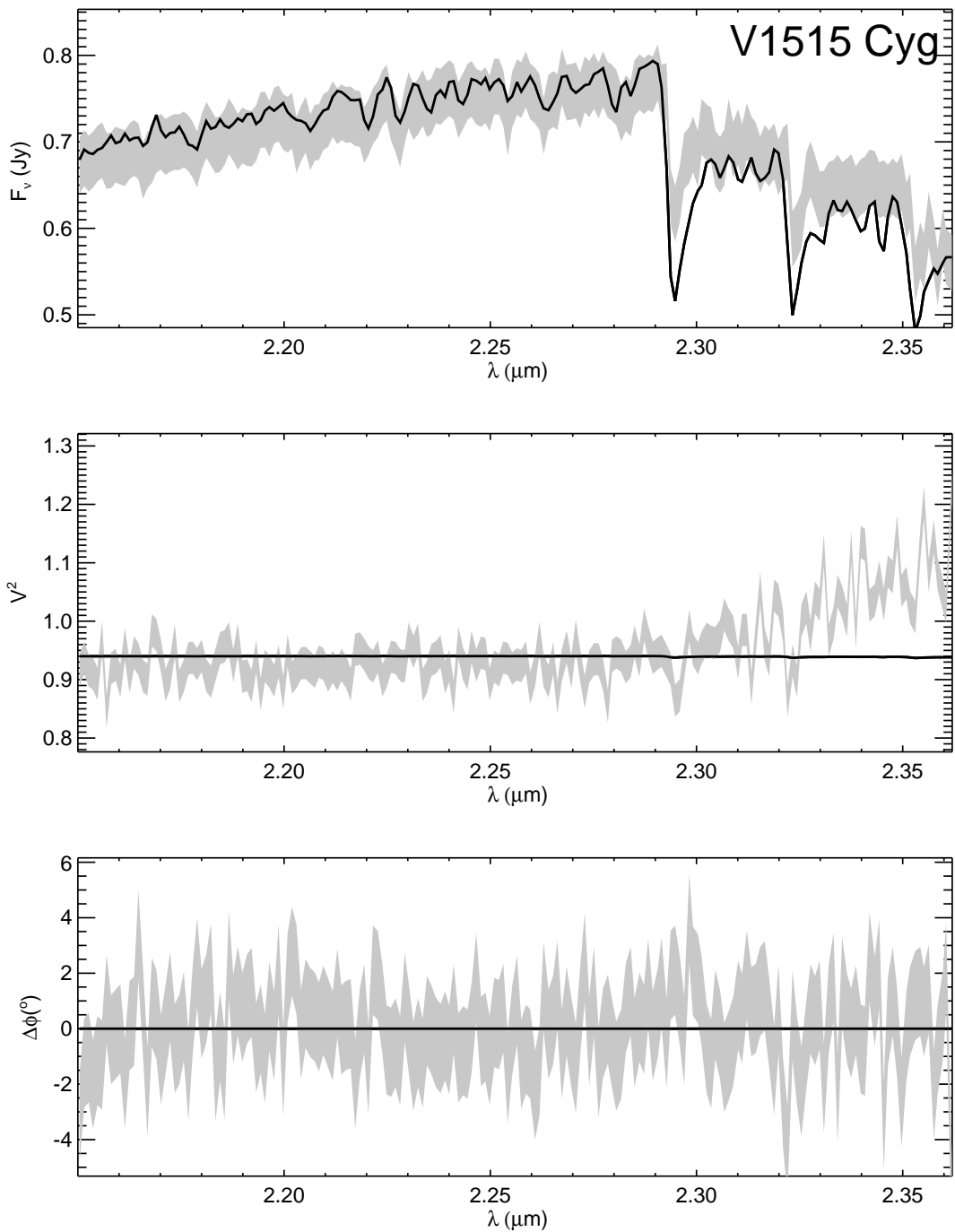}
\caption{Fluxes, squared visibilities, and differential phases
  observed for V1515 Cyg (data and 1$\sigma$ uncertainties shown with
  gray shaded regions).  We also plot the fluxes, $V^2$, and
  $\Delta \phi$ predicted by our best-fit disk model ({\it solid
    curves}).  Note that the
  $V^2$ values greater than unity, seen for $\lambda \ga 2.33$ $\mu$m,
  are unphysical and are probably caused by lower signal-to-noise near
  the band edges.
\label{fig:v1515cyg}}
\end{figure}

We measured fluxes, squared visibilities ($V^2$), and differential
phases ($\Delta \phi$) for our targets and calibrator stars in each of
the 330 spectral channels across the $K$-band provided by the grism
(Figures \ref{fig:fuori}--\ref{fig:v1515cyg}). 
The data calibration is described in detail in \citet{EISNER+10}, but
we provide a basic sketch here.  

Normalized fluxes are calculated by
dividing observed source and calibrator spectra and then multiplying
by an appropriate stellar template for the calibrator \citep[given by a
Nextgen model atmosphere;][]{HAB99}.  The absolute flux level and
slope of the spectra are then matched to near-IR photometry from 2MASS.
Error bars for the calibrated fluxes reflect channel-to-channel
uncertainties, but do not include uncertainties in the overall normalization.

Squared visibilities for targets are divided by $V^2$ measured for
calibrators.  As for the fluxes, error bars reflect only
channel-to-channel uncertainties.  We do not include the uncertainty
in the overall normalization of the $V^2$.

Differential phases--i.e., the phase in each spectral
channel relative to some fiducial channel--are calibrated by
subtracting the $\Delta \phi$  measured for the calibrator stars.  
The broadband slope and curvature of the measured differential phases
are altered by instrumental effects.  The slope is altered by errors
in group delay tracking, and curvature in the phase versus wavelength
may be introduced by atmospheric water vapor turbulence and
dispersion and refraction during internal beam-transport \citep[see][]{POTT+10}.
Hence we can not attribute any broadband 
slope or curvature in the data to the
observed source.  We therefore remove the slope and curvature seen
across our bandpass from the
differential phase data, as described in \citet{POTT+10}.

\section{General Features of the Data \label{sec:general}}
All objects in our sample show strong absorption in the spectral regions
corresponding to the CO ro-vibrational bandheads, $v=2\rightarrow 0$, $3
\rightarrow 1$, and $4 \rightarrow 2$ (Figures
\ref{fig:fuori}--\ref{fig:v1515cyg}).  The rest wavelengths of these three
bandheads are 2.2936, 2.3227, and 2.3527 $\mu$m, respectively.

FU Ori and V1057 Cyg also show dips
in $V^2$ at the wavelengths corresponding to these spectral features.
V1515 Cyg may show similar dips relative to the $V^2$ seen in adjacent
channels corresponding to continuum emission, but these dips are not
significant given the noise in these data.

Lower $V^2$ generally means more resolved emission.  However, because
the CO appears in absorption, the dips in $V^2$ actually indicate that the CO
distribution is more compact than the continuum emission.
This can be understood by considering a compact, hot region of the disk
that includes CO absorption, and a cooler, continuum-emitting region at
larger radii.  At wavelengths where CO absorbs emission from the hot,
compact component, the flux-weighted size of the two regions moves to
larger stellocentric radii.  Thus, the compact CO absorption leads to a
larger measured size at these wavelengths, and hence a dip in $V^2$.

The observed differential phases indicate a centroid offset between
the continuum emission and the CO absorbing material.  This offset is
most obvious for FU Ori, and marginally significant for V1057
Cyg.   For V1515 Cyg the data are too noisy to argue the presence or
absence of such an offset.  Differential phase offsets in the CO
bandheads indicate that the annuli of the disk that
have appropriate temperatures for CO absorption in their spectra must have
centroids that differ from annuli that are too
hot for CO to exist or cool enough that they produce continuum
emission.

The spectra for all objects show an absorption feature at the central
wavelength of Br$\gamma$, 2.1662 $\mu$m.  In addition, the spectra
show hints of some emission on the red side of the line.  These
spectral shapes are consistent with P Cygni profiles seen in other
hydrogen lines \citep[e.g.,][]{HPD03}.

One object--V1057 Cyg--also shows a significant dip in $V^2$ in the
Br$\gamma$ spectral region.  Since the line is predominantly seen in
absorption, the same argument applied to CO above suggests that the
Br$\gamma$ absorption occurs in more compact regions than the
continuum emission.  The Br$\gamma$ feature is more prominent in the
spectrum of V1057 Cyg than in FU Ori or V1515 Cyg; the higher
line-to-continuum ratio of Br$\gamma$ in this object may explain why
the $V^2$ dip is only seen here.

The CO bandhead features are much broader in V1057 Cyg than in other
objects in our sample.  These broad features do not appear to be due
to instrumental effects.  During the night of our V1057 Cyg
observations, we observed known CO-emitting objects immediately before
and after V1057 Cyg.  Neither of these objects show such broad
features \citep[see][]{EISNER+10c}.  We also verified that the
appearance of the CO features in V1057 Cyg persists across multiple
observations and for different calibrator stars.
Thus the broadening seen in
the CO lines of V1057 Cyg appears to be intrinsic to the source.

\section{Modeling \label{sec:mod}}
In this section we attempt to
create a physically-motivated cirumstellar disk model to explain the
measured fluxes, squared visibilities, and differential phases.
In our modeling we assume that the emission from our targets arises
entirely from circumstellar material.  In previous modeling of FU Ori
sources spectrophotometric data has been fitted well with no
contribution from a compact stellar photosphere
\citep[e.g.,][]{ZHU+08}.    We therefore do not require any
correction for unresolved flux from the central star.  

We begin with a geometrically thin disk model.  We assume the disk
spans a range of radii from $R_{\rm in}$ to $R_{\rm out}$, and has a
temperature profile $T = T_{\rm in} (R/R_{\rm in})^{\alpha}$.  
We divide the disk up into a series of annuli and assume the emergent
spectrum of each annulus is that of a supergiant star.  
For FU Ori stars this is a reasonable assumption since they are heated primarily
from accretion energy released near the midplane.  Disk atmosphere
models have been constructed specifically for FU Ori systems, and
these do provide better fits to detailed spectrophotometric data than
models based on stellar atmospheres \citep[e.g.,][]{HHC04}.  We are,
however, not concerned with this level of detail here, and we deem our
models based on giant star atmospheres to be sufficient.

For annular temperatures between 1700 and 10000 K, we use emergent
spectra for stars with $\log g = 0$ \citep{HAB99}. (In practice, the
maximum temperature is typically between 4000-6000 K.)  For $T<1700$ K, we
assume that the emission is a simple blackbody.  The emergent spectrum
is rotationally broadened by the Keplerian velocity expected for each
annulus, $v \sin i = \sqrt{G M / R} \sin i$.  We take $M = 0.5$
M$_{\odot}$ for our objects; this is compatible with the dynamical
mass of $0.3 \pm 0.1$ M$_{\odot}$ for FU Ori determined by
\citet{ZHU+09} and the pre-outburst spectral type of V1057 Cyg.
The spectrum of the disk model is given by the sum of fluxes over all
annuli.

The visibility of each annulus is given by the visibility of a uniform
ring \citep[e.g.,][]{EISNER+04}.  The visibilities for the entire disk
are computed as the flux-weighted average of the annular visibilities:
\begin{equation}
V^2_{\rm disk} = \left(\frac{\sum F_{\rm annulus} V_{\rm annulus}}{\sum
  F_{\rm annulus}}\right)^2.
\end{equation}

The differential phase is related to the centroid offset.  For each
annulus, the differential phase is 
\begin{equation}
\Delta \phi_{\rm annulus} \approx {2 \pi r_{\rm uv}} \Delta
\theta_{\rm annulus},
\end{equation}
where $r_{\rm uv}$ is the $uv$ radius \citep[see,
e.g.,][]{EISNER+04} and $\Delta \theta_{\rm annulus}$ is the centroid
offset.  This approximation is valid as long as the emission is not
well-resolved at the angular resolution of the observations.
The differential phase signal for the entire model is \citep[see,
e.g.,][]{EISNER+10}
\begin{equation}
\Delta \phi_{\rm disk} = \frac{\sum \Delta \phi_{\rm annulus} V_{\rm
    annulus} F_{\rm annulus}}{\sum V_{\rm annulus} F_{\rm annulus}}.
\end{equation}
As described above, the slope and curvature of the observed, broadband
differential phase are altered by instrumental effects.
We therefore remove broadband slope and curvature from both
observed and modeled $\Delta \phi$.

The simple, geometrically thin disk model considered above would not
produce centroid offsets.  However if we allowed the disk to be
somewhat asymmetric, certain disk radii could have centroids offset from
those of other disk radii.  We allow for the
possibility of such asymmetry by including a centroid offset that changes
linearly with disk radius.  This aspect of the model is described by a
single free parameter that gives the amplitude of this offset.  We denote
this amplitude by $\Delta \theta_{\rm max}$.

To account for the possibility that some of the near-IR emission
arises from more spatially extended regions--for example in remnant
envelopes \citep[as implied
by the results of][]{MILLAN-GABET+06b}--we include a scattered light
contribution in our models.  While there may also be thermal emission
from a more extended component, the cooler temperatures at large
stellocentric radii would produce a very small fractional flux in the near-IR.
We assume that the extended, scattered emission has the same
spectrum as the disk, and is sufficiently extended so that is is
completely resolved ($V^2=0$) in our observations.  Writing the
fractional scattered flux, $f=F_{\rm scat} / F_{\rm disk}$, the model
fluxes and visibilities are modified to
\begin{equation}
F_{\rm model} = F_{\rm disk} (1+f),
\end{equation}
\begin{equation}
V^2_{\rm model} = V^2_{\rm disk} (1+f)^{-2}.
\end{equation}

Finally, we include two scaling factors in the model that take into
account additional uncertainties not reflected in the error bars for
the data.  The first is a constant scaling in the overall flux level,
$X_{\rm flux}$.  This scaling accounts for uncertainties in the flux
calibration beyond the assumed channel-to-channel errors, including
uncertainties in the 2MASS photometry used to normalize the
spectra, or source dimming between the photometric and interferometric
observations.  We also include a constant scaling in the overall $V^2$
normalization, $X_{V^2}$, to account for overall calibration uncertainties beyond
the channel-to-channel uncertainties reflected in the plotted error
bars.  These two scale factors may also absorb uncertainty in the
assumed source distances (Table \ref{tab:sample}).  It turns out that
these scaling factors are near unity in our best-fit models, and hence
are not crucial to the fitting process; we include them
for completeness.

For FU Ori, the disk inclination and position angle have been measured
using previous interferometric observations.  In the near-IR the 
inclination and position angle were estimated as $\sim 55^{\circ}$ and
$47^{\circ}$, respectively \citep{MALBET+05}.  In the mid-IR, the
inclination and position angle were found to be $\sim 55^{\circ}$ and
$100^{\circ}$ \citep{QUANZ+06}.  We adopt the near-IR-determined
values here.  For V1057 Cyg and V1515 Cyg, no such geometrical
measurements are available, and so we leave inclination and position
angle as free parameters in our models.  We can not, however,
constrain these well with the limited $uv$ coverage of our
observations.

Given the large numbers of free parameters, we fit our models to the
data using a non-linear gradient-fitting algorithm based on the
Levenberg-Marquardt method.  Such algorithms may fall into local
minima in the $\chi^2$ surface, and so initial parameter guesses and
step sizes for gradient searches influence the results.  We have
selected step sizes that ensure that all free
parameters are varied during the fitting process.  
However we are hesitant to claim
that the``best-fit'' models we determine are the true, global best fits.

To improve the signal-to-noise ratio in the fitting, we only use data
with wavelength between 2.15 and 2.36 $\mu$m to constrain the models.
Data at the band edges are substantially noisier because of instrumental
vignetting and contributions from ambient water vapor
(Pott et al. 2010; Woillez et al. 2011).  

\section{Modeling Results \label{sec:models}}

\subsection{FU Ori}

\begin{figure}
\plotone{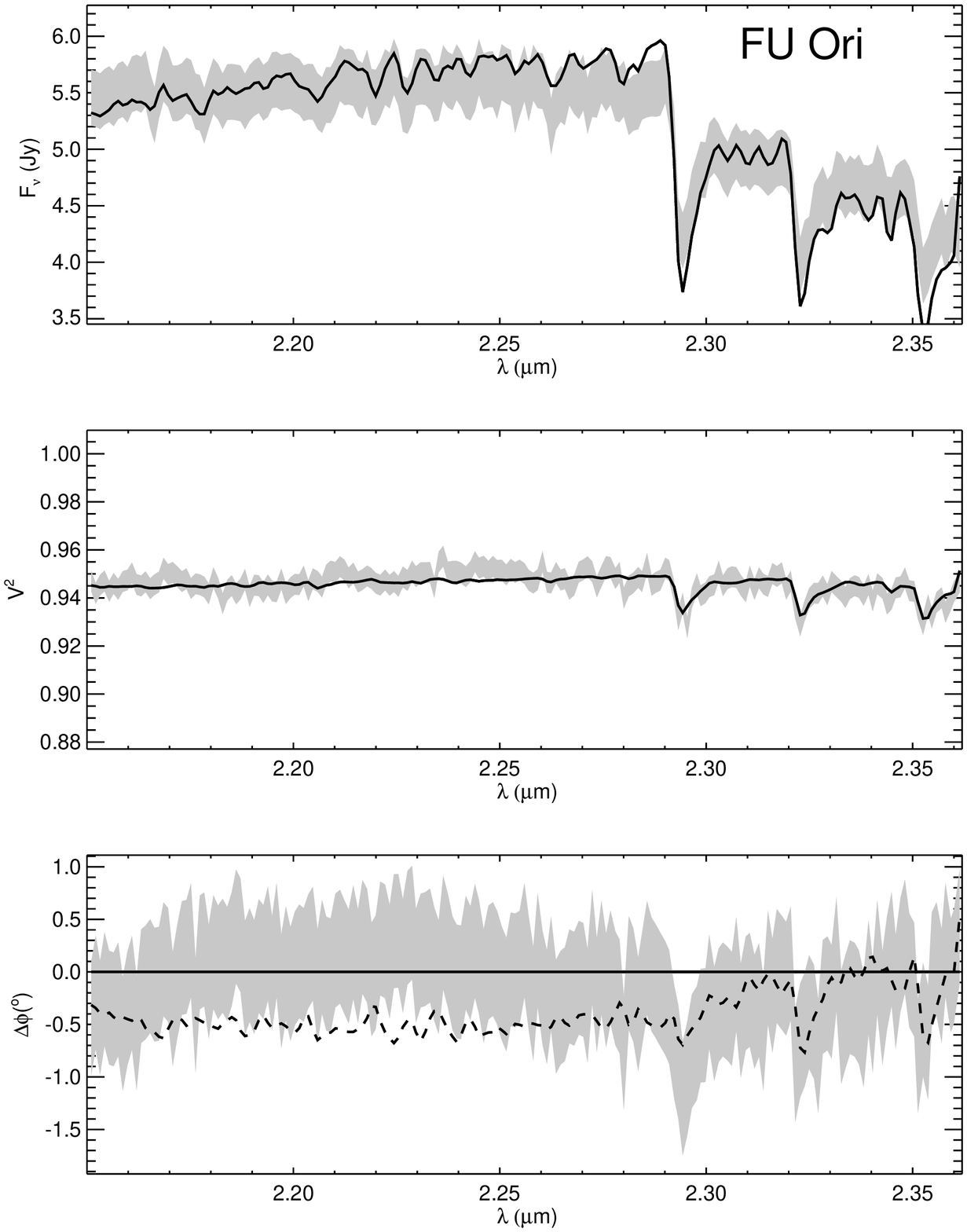}
\caption{Fluxes, squared visibilities, and differential phases
  observed for FU Ori (data and 1$\sigma$ uncertainties shown with
  gray shaded regions).  We also plot the fluxes, $V^2$, and
  $\Delta \phi$ predicted by our best-fit disk model ({\it solid
    curves}).  The best-fit model predicts a flat differential
  phase across the observed bandpass, and misses the significant
  deviations in $\Delta \phi$ seen in and out of the CO bandheads.  
  A model exhibiting a  centroid offset with $\Delta
  \theta_{\rm max} = 0.8$ mas ({\it dashed curve})
  fits the differential phases in the CO bandhead region
  well. 
 \label{fig:fuori}}
\end{figure}

Our best-fit disk model for FU Ori is shown in Figure
\ref{fig:fuori}.  The model fits quite well in this wavelength
region (reduced $\chi^2 = 0.5$).  
While the modeled differential phases do not match the
observations perfectly, this is likely due to imperfectly subtracted
slope and curvature (which are instrumental effects).  Indeed,
residual curvature of the differential phase data forces the best-fit
model toward a flat differential phase versus wavelength.

If we consider the differential phase data only over the narrower
spectral region containing the CO
bandheads (where broadband slope and curvature effects are less important) we
can easily construct a model that agrees with the data well.  In
particular, we use the best-fit model parameters for FU Ori listed in Table
\ref{tab:results}, increase the value of $\theta_{\rm max}$, and
subtract a constant phase offset.  Note that $\theta_{\rm max}$ has no
impact on the fluxes or $V^2$ values and so we are free to modify its
value in order to better-fit the differential phases.  The dashed line
in Figure \ref{fig:fuori} shows that a model with a large centroid
offset, $\theta_{\rm max}= 0.8$ mas, produces a differential phase
signature that agrees well with the observations in the CO bandhead
region.

The differential phase signature seen for FU Ori thus can not
be explained with a simple, geometrically thin disk model.  Rather we
require a centroid offset that depends on stellocentric radius.  A
disk warp can obscure part of the outer regions of a disk while
leaving the inner regions unobscured (or vice versa).  The centroid
offsets implied by our data suggest that the FU Ori inner disk may be
significantly warped.  In our model this warp occurs over
stellocentric radii spanning 0.07 AU to 0.49 AU.

Disk winds may also cause centroid offsets that
vary as a function of stellocentric radius
\citep[e.g.,][]{CHK93,HC95}.  Since all of the sources in our sample
are known to be losing mass via winds \citep[e.g.,][]{CHA87}, this may be
an alternative explanation for the differential phase signatures seen in
the data.  In any event, the large differential phases observed across
the spectral region including CO bandhead absorption indicate 
substantial geometric asymmetry in the underlying emission
from FU Ori.

The outer radius derived for FU Ori is $\sim 0.5$ AU, similar to the
$\sim 0.5$--1 AU value derived in the the SED modeling of
\citet{ZHU+07}.  This likely represents the outer radius of the hot,
high-accretion rate, inner disk component.  As argued in
\citet{ZHU+07}, this implies that the high accretion rate in FU Ori is
probably not caused by pure thermal instability, but rather
gravitational or MRI instability.  Note that a larger $R_{\rm out}$
value is not required to fit the broadband SED, and our model fits
photometry out to $\lambda \approx 10$ $\mu$m well.  To fit longer
wavelength photometry requires an additional model component such as a
flared disk \citep[e.g.,][]{ZHU+08}.

\subsection{V1057 Cyg}

\begin{figure}
\plotone{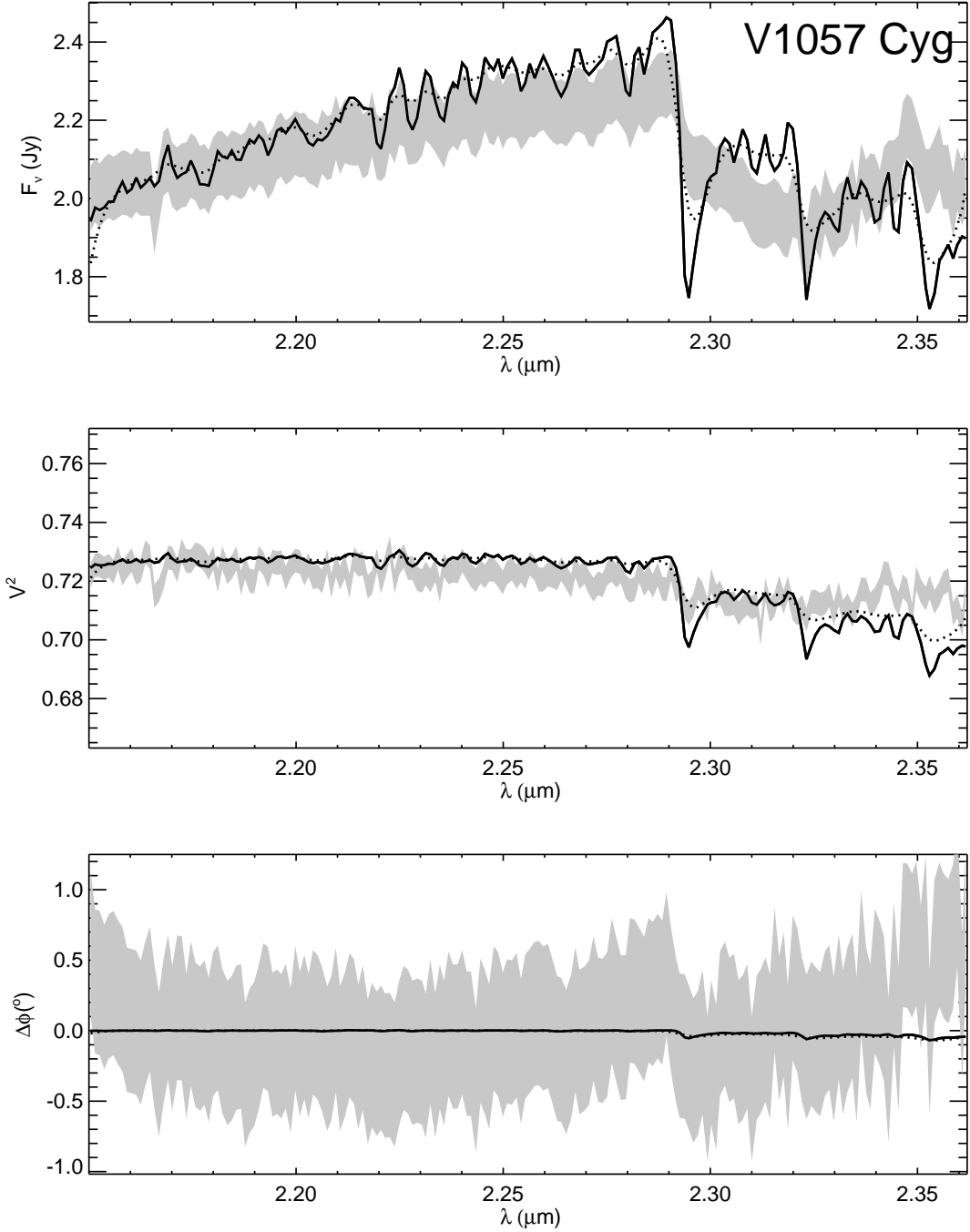}
\caption{Fluxes, squared visibilities, and differential phases
  observed for V1057 Cyg (data and 1$\sigma$ uncertainties shown with
  gray shaded regions).  We plot the fluxes, $V^2$, and
  $\Delta \phi$ predicted by our best-fit disk model  with solid curves. 
We also show the predictions of a model where we scaled the velocities
up so that the CO lines are rotationally broadened  by $\sim 500$ km
s$^{-1}${\it (dotted curves)} to illustrate how a disk wind might
appear different than a rotating disk.  The model with higher
velocities produces a reduced $\chi^2$ of 1.3, compared to 1.9 for the
baseline disk model.
\label{fig:v1057cyg}}
\end{figure}

For V1057 Cyg the disk model does not fit the CO bandhead features
well (Figure \ref{fig:v1057cyg}).   The reduced $\chi^2$ residuals of
the fit to our $K$-band data  are 1.9.  The discrepancy is due to the
extremely broad CO overtone absorption features.  For
an assumed 0.5 M$_{\odot}$ central object, the Keplerian velocities do
not exceed $\sim 100$ km s$^{-1}$, and hence the modeled CO
bandheads are not broadened significantly.  

A higher inclination would yield higher $v \sin i$ values, but would
also lower the modeled flux and make it difficult to fit the
observations.  Moreover even for $i=90^{\circ}$ the synthetic spectra
are still not as broad as the data.  A higher central mass can also
increase the Keplerian velocities. However $M_{\ast} > 50$ M$_{\odot}$ is
required to produce the $\sim 500$ km s$^{-1}$ broadening seen in the
data.  A stellar mass this high is not (remotely) compatible with
the pre-outburst, K-spectral type determined for V1057 Cyg
\citep{HERBIG58}.  

Such high velocities could, however, be created by a
rotating disk wind \citep[e.g.,][]{CHK93}.
V1057 Cyg is known to possess a variable, high-velocity wind
\citep{HERBIG09,CHA87}.   If this wind dominated the near-IR absorption
during the epoch of our observations, it may explain the broad CO
absorption features in our data.  While previous observations of V1057
Cyg found the CO bandheads to be narrower than inferred here
\citep[][]{CHL87,HK87},  the CO overtone absorption spectra show
substantial variability \citep{BISCAYA+97}. 
When the wind is more dominant, the accretion luminosity generated in
the disk midplane may be absorbed by CO (and other matter) 
with radial velocities substantially higher than Keplerian.

Rotating disk-wind models are substantially more complex than the
simple disk models considered here.  Rather than creating an accurate
wind model, we demonstrate the overall features of such a model with a
simple modification of our baseline disk model.  Specifically, we
scale up the Keplerian rotation profile to account for the possibility
of higher radial velocities due to outflowing motions.

Figure \ref{fig:v1057cyg} shows that a model with substantially higher
radial velocities does indeed fit the data better than the simple disk
model.  Here we scaled the disk velocity structure up (arbitrarily) by
a factor of 30, so that the CO lines are broadened by $v\sin i \sim
500$ km s$^{-1}$.  This model fits the data substantially better than
the Keplerian disk model considered above; the high-velocity model
produces a reduced $\chi^2$ value of 1.3.
While the fit deviates from the data for the CO $v=4\rightarrow 2$
transition, this deviation may be due, in part, to lower signal-to-noise 
of the data toward the band edge.

Our best-fit disk model also fails to reproduce the $V^2$ dip
in the Br$\gamma$ spectral region observed for V1057 Cyg.  Because the
inner disk temperature is $\approx 4000$ K for our best-fit disk
model, there is essentially no Br$\gamma$ absorption in the model
spectra.  To explain the Br$\gamma$ feature seen in the data would
require an additional model component with a higher temperature.
This component would also have to be reside at very small
stellocentric radii ($\la 0.1$ AU, which is more compact than the
distribution of CO-absorbing material) to produce the observed dip in $V^2$.
Across the K-band, a simple disk model is not an obvious match to our
observations of V1057 Cyg.

\subsection{V1515 Cyg}

The $V^2$ and $\Delta \phi$ observed for V1515 Cyg are noisy compared
to the data for the other two objects in our sample.  V1515 Cyg is
more than a magnitude fainter than V1057 Cyg, and $\sim 3$ mags
fainter than FU Ori, and so more noise is expected. V1515
Cyg is near (but within) the brightness limit of the SPR mode, beyond
which signal-to-noise degrades significantly (Woillez et al. 2011).
However, lower throughput near the band edges means that the
signal-to-noise is degraded more in the spectral region including CO.
This low signal-to-noise gives rise to the un-physically large $V^2$
(i.e., normalized $V^2>1$) seen for $\lambda \ga 2.33$ $\mu$m.

The fluxes for V1515 Cyg appear reasonably consistent with our simple
disk model (Figure \ref{fig:v1515cyg}).  The $V^2$ and $\Delta \phi$ are
not very consistent with the data, as seen in the high reduced
$\chi^2$ value for the fit (Table \ref{tab:results}).  The large
$\chi^2$ value is due, at least in part, to underestimated
uncertainties.  The large $V^2$ seen at the long-wavelength edge of
the band are unphysical, and so the plotted error bars are clearly
underestimated.  Over most of the band the model appears compatible
with the noisy $V^2$ and $\Delta \phi$ data.

\section{Conclusions}
We presented near-infrared, spatially resolved, spectroscopic
observations of FU Ori, V1057 Cyg, and V1515 Cyg, obtained with the
Keck Interferometer.  With $R=2000$ across the $K$-band, we spectrally
resolved individual CO bandheads as well as the Br$\gamma$ transition
of hydrogen.
Our measured fluxes, squared visibilities, and differential phases
were compared with simple accretion disk models.  

A disk model is consistent with the data for FU Ori, although we
require some asymmetry in the inner $\sim 1$ AU of the 
disk to explain the measured,
non-zero, differential phases.  We speculate that the inner disk of FU Ori may
be warped. The disk model also appears reasonably compatible with the
data for V1515 Cyg, although these data are noisy and so the
constraints are weak.   

For V1057 Cyg the disk model does not fit our data well. 
The observed CO bandhead features are broader than expected
for a disk in Keplerian rotation.  The disk model also fails to reproduce
compact Br$\gamma$ emission seen in the data.
We speculate that a rotating, high-velocity, inner disk wind may be
contributing substantially to the observed near-IR emission in this
source, at least at the epoch of our observations.

Overall this study suggests a more complex explanation for the near-IR
observations of three canonical FU Ori stars than the unified picture of
a simple accretion disk model.  In addition to a rotating disk
undergoing viscous accretion, outflow is also likely to be an
important element in the interpretation of radiative and kinematic
data.  While the namesake object, FU Ori, is consistent with a
simple accretion disk model, V1057 Cyg is not.  It is possible that
V1057 Cyg is undergoing an epoch of enhanced outflow, and that other
FU Ori objects experience similar episodes.  However further
observations, at additional epochs, are needed to test this hypothesis.

\vspace{0.2in}
Data presented herein were obtained at the W. M. Keck Observatory,
from telescope time allocated to the National Aeronautics and Space 
Administration through the agency's scientific partnership with the California 
Institute of Technology and the University of California. The Observatory was 
made possible by the generous financial support of the W. M. Keck Foundation. 
The authors wish to recognize and acknowledge the cultural role and reverence
that the summit of Mauna Kea has always had within the indigenous Hawaiian
community. We are most fortunate to have the opportunity to conduct 
observations from this mountain. The ASTRA program, which enabled the
observations presented here, was made possible
by funding from the NSF MRI grant AST-0619965.  
The Keck Interferometer is funded by the National Aeronautics and
Space Administration as  part of its Exoplanet Exploration program.
This work has used software from 
NExSci at the California Institute of Technology.
We are grateful to the
ASTRA team for their support of the new instrumental mode used here, and for their 
enthusiasm for the science enabled.  We also thank
R. Akeson, A. Ghez, R. Millan-Gabet, J. Monnier, and J.-U. Pott 
for useful comments on the manuscript.

\epsscale{0.85}

\epsscale{0.85}

\epsscale{0.85}

\clearpage
\begin{deluxetable}{lcccccccc}
\tabletypesize{\scriptsize}
\tablewidth{0pt}
\tablecaption{Targets and Observations
\label{tab:sample}}
\tablehead{\colhead{Source} & \colhead{$\alpha$} 
& \colhead{$\delta$} & \colhead{$d$} & \colhead{Observation Date}  &
\colhead{$B_{\rm proj}$} & \colhead{$B_{\rm PA}$} & \colhead{Calibrators} \\
 & (J2000) & (J2000) & (pc) & (UT) & (m) & ($^{\circ}$) &}
\startdata
FU Ori & 05 45 22.36 & +09 04 12.30 & 412 &  25 Nov 2010 & 82-84 &
53-55 & HD 37147, HD 42807 \\
V1057 Cyg & 20 58 53.73 & +44 15 28.54 & 600 & 15 July 2009 & 
78-81 & 46-53 & HD 192985, HD 195050 \\
V1515 Cyg & 20 23 48.02 &  +42 12 25.80 & 1000 & 21 July 2010 & 
78-80 & 45-48 & HD 194206, HIP 102667
\enddata
\tablecomments{$B_{\rm proj}$ and $B_{\rm PA}$ are the projected length and
  position angle of the KI baseline on the sky during our
  observations.  The distance to FU Ori is based on trigonometric
  parallax measured to the Orion region \citep{REID+09}.  V1057 Cyg
  lies in the North American Nebula, for which the distance has been
  estimated to be 600 pc \citep{LS02}.  V1515 Cyg
  resides in the Cyg R1 region, for which the distance has been
  estimated as 1000 pc \citep{RACINE68}.}
\end{deluxetable}

\begin{deluxetable}{lccc}
\tabletypesize{\small}
\tablewidth{0pt}
\tablecaption{Disk Model Fits
\label{tab:results}}
\tablehead{\colhead{$ $} & \colhead{FU Ori} & \colhead{V1057 Cyg} &
  \colhead{V1515 Cyg}}
\startdata
$\chi_{\rm r}^2$ & 0.5 & 1.9 & 2.6 \\
$R_{\rm in}$ (AU) &    0.07$\pm$   0.02 &    0.03$\pm$   0.07 &    0.03$\pm$   0.11 \\
$R_{\rm out}$ (AU) &    0.49$\pm$   0.03 &    3.54$\pm$   3.14 &    1.02$\pm$   0.20 \\
$\alpha$ &   -0.95$\pm$   0.10 &   -0.50$\pm$   0.03 &   -0.72$\pm$   0.13 \\
$T_{\rm in}$ (K) &  5529$\pm$  779 &  4054$\pm$ 1169 &  5973$\pm$ 1548 \\
$f$ &    0.02$\pm$0.01 &    0.11$\pm$0.01 &    0.03$\pm$0.03 \\
$\Delta \theta$ (mas) &    0$\pm$0 &    36$\pm$  104 &    0$\pm$ 0 \\
$X_{V^2}$ &    1.1$\pm$0.1 &    1.0$\pm$0.1 &    1.0$\pm$1.0 \\
$X_{\rm flux}$ &    1.0$\pm$0.1 &    1.0$\pm$0.1 &    1.0$\pm$1.0 \\
\enddata
\tablecomments{Parameters are defined in \S \ref{sec:mod}.  While we
  also fitted inclination and position angle for V1057 Cyg and V1515
  Cyg, these very poorly constrained by our data and so we have not
  listed the fitted values here.  
}
\end{deluxetable}


\begin{thebibliography}{40}
\expandafter\ifx\csname natexlab\endcsname\relax\def\natexlab#1{#1}\fi

\bibitem[{{Armitage} {et~al.}(2001){Armitage}, {Livio}, \& {Pringle}}]{ALP01}
{Armitage}, P.~J., {Livio}, M., \& {Pringle}, J.~E. 2001, \mnras, 324, 705

\bibitem[{{Bell} \& {Lin}(1994)}]{BL94}
{Bell}, K.~R. \& {Lin}, D.~N.~C. 1994, \apj, 427, 987

\bibitem[{{Biscaya} {et~al.}(1997){Biscaya}, {Rieke}, {Narayanan}, {Luhman}, \&
  {Young}}]{BISCAYA+97}
{Biscaya}, A.~M., {Rieke}, G.~H., {Narayanan}, G., {Luhman}, K.~L., \& {Young},
  E.~T. 1997, \apj, 491, 359

\bibitem[{{Calvet} {et~al.}(1993){Calvet}, {Hartmann}, \& {Kenyon}}]{CHK93}
{Calvet}, N., {Hartmann}, L., \& {Kenyon}, S.~J. 1993, \apj, 402, 623

\bibitem[{{Carr} {et~al.}(1987){Carr}, {Harvey}, \& {Lester}}]{CHL87}
{Carr}, J.~S., {Harvey}, P.~M., \& {Lester}, D.~F. 1987, \apjl, 321, L71

\bibitem[{{Clarke} {et~al.}(2005){Clarke}, {Lodato}, {Melnikov}, \&
  {Ibrahimov}}]{CLARKE+05}
{Clarke}, C., {Lodato}, G., {Melnikov}, S.~Y., \& {Ibrahimov}, M.~A. 2005,
  \mnras, 361, 942

\bibitem[{{Colavita} {et~al.}(2003){Colavita}, {Akeson}, {Wizinowich}, {Shao},
  {Acton}, {Beletic}, {Bell}, {Berlin}, {Boden}, {Booth}, {Boutell}, {Chaffee},
  {Chan}, {Chock}, {Cohen}, {Crawford}, {Creech-Eakman}, {Eychaner},
  {Felizardo}, {Gathright}, {Hardy}, {Henderson}, {Herstein}, {Hess},
  {Hovland}, {Hrynevych}, {Johnson}, {Kelley}, {Kendrick}, {Koresko}, {Kurpis},
  {Le Mignant}, {Lewis}, {Ligon}, {Lupton}, {McBride}, {Mennesson},
  {Millan-Gabet}, {Monnier}, {Moore}, {Nance}, {Neyman}, {Niessner}, {Palmer},
  {Reder}, {Rudeen}, {Saloga}, {Sargent}, {Serabyn}, {Smythe}, {Stomski},
  {Summers}, {Swain}, {Swanson}, {Thompson}, {Tsubota}, {Tumminello}, {van
  Belle}, {Vasisht}, {Vause}, {Walker}, {Wallace}, \& {Wehmeier}}]{COLAVITA+03}
{Colavita}, M., {Akeson}, R., {Wizinowich}, P., {Shao}, M., {Acton}, S.,
  {Beletic}, J., {Bell}, J., {Berlin}, J., {Boden}, A., {Booth}, A., {Boutell},
  R., {Chaffee}, F., {Chan}, D., {Chock}, J., {Cohen}, R., {Crawford}, S.,
  {Creech-Eakman}, M., {Eychaner}, G., {Felizardo}, C., {Gathright}, J.,
  {Hardy}, G., {Henderson}, H., {Herstein}, J., {Hess}, M., {Hovland}, E.,
  {Hrynevych}, M., {Johnson}, R., {Kelley}, J., {Kendrick}, R., {Koresko}, C.,
  {Kurpis}, P., {Le Mignant}, D., {Lewis}, H., {Ligon}, E., {Lupton}, W.,
  {McBride}, D., {Mennesson}, B., {Millan-Gabet}, R., {Monnier}, J., {Moore},
  J., {Nance}, C., {Neyman}, C., {Niessner}, A., {Palmer}, D., {Reder}, L.,
  {Rudeen}, A., {Saloga}, T., {Sargent}, A., {Serabyn}, E., {Smythe}, R.,
  {Stomski}, P., {Summers}, K., {Swain}, M., {Swanson}, P., {Thompson}, R.,
  {Tsubota}, K., {Tumminello}, A., {van Belle}, G., {Vasisht}, G., {Vause}, J.,
  {Walker}, J., {Wallace}, K., \& {Wehmeier}, U. 2003, \apjl, 592, L83

\bibitem[{{Colavita} \& {Wizinowich}(2003)}]{CW03}
{Colavita}, M.~M. \& {Wizinowich}, P.~L. 2003, in Interferometry for Optical
  Astronomy II. Edited by Wesley A. Traub. Proceedings of the SPIE, Volume
  4838, pp. 79-88 (2003)., 79--88

\bibitem[{{Croswell} {et~al.}(1987){Croswell}, {Hartmann}, \& {Avrett}}]{CHA87}
{Croswell}, K., {Hartmann}, L., \& {Avrett}, E.~H. 1987, \apj, 312, 227

\bibitem[{{Eisner} {et~al.}(2010{\natexlab{a}}){Eisner}, {Akeson}, {Colavita},
  {Ghez}, {Graham}, {Hillenbrand}, {Millan-Gabet}, {Monnier}, {Pott},
  {Ragland}, {Wizinowich}, \& {Woillez}}]{EISNER+10c}
{Eisner}, J.~A., {Akeson}, R., {Colavita}, M., {Ghez}, A., {Graham}, J.,
  {Hillenbrand}, L., {Millan-Gabet}, R., {Monnier}, J.~D., {Pott}, J.~U.,
  {Ragland}, S., {Wizinowich}, P., \& {Woillez}, J. 2010{\natexlab{a}}, in
  Society of Photo-Optical Instrumentation Engineers (SPIE) Conference Series,
  Vol. 7734, Society of Photo-Optical Instrumentation Engineers (SPIE)
  Conference Series

\bibitem[{{Eisner} {et~al.}(2004){Eisner}, {Lane}, {Hillenbrand}, {Akeson}, \&
  {Sargent}}]{EISNER+04}
{Eisner}, J.~A., {Lane}, B.~F., {Hillenbrand}, L., {Akeson}, R., \& {Sargent},
  A. 2004, \apj, 613, 1049

\bibitem[{{Eisner} {et~al.}(2010{\natexlab{b}}){Eisner}, {Monnier}, {Woillez},
  {Akeson}, {Millan-Gabet}, {Graham}, {Hillenbrand}, {Pott}, {Ragland}, \&
  {Wizinowich}}]{EISNER+10}
{Eisner}, J.~A., {Monnier}, J.~D., {Woillez}, J., {Akeson}, R.~L.,
  {Millan-Gabet}, R., {Graham}, J.~R., {Hillenbrand}, L.~A., {Pott}, J.,
  {Ragland}, S., \& {Wizinowich}, P. 2010{\natexlab{b}}, \apj, 718, 774

\bibitem[{{Green} {et~al.}(2006){Green}, {Hartmann}, {Calvet}, {Watson},
  {Ibrahimov}, {Furlan}, {Sargent}, \& {Forrest}}]{GREEN+06}
{Green}, J.~D., {Hartmann}, L., {Calvet}, N., {Watson}, D.~M., {Ibrahimov}, M.,
  {Furlan}, E., {Sargent}, B., \& {Forrest}, W.~J. 2006, \apj, 648, 1099

\bibitem[{{Greene} {et~al.}(2008){Greene}, {Aspin}, \& {Reipurth}}]{GAR08}
{Greene}, T.~P., {Aspin}, C., \& {Reipurth}, B. 2008, \aj, 135, 1421

\bibitem[{{Hartmann} \& {Calvet}(1995)}]{HC95}
{Hartmann}, L. \& {Calvet}, N. 1995, \aj, 109, 1846

\bibitem[{{Hartmann} {et~al.}(2004){Hartmann}, {Hinkle}, \& {Calvet}}]{HHC04}
{Hartmann}, L., {Hinkle}, K., \& {Calvet}, N. 2004, \apj, 609, 906

\bibitem[Hartmann \& Kenyon(1996)]{HK96} 
Hartmann, L., \& Kenyon, S.~J.\ 1996, \araa, 34, 207 

\bibitem[{{Hartmann} \& {Kenyon}(1985)}]{HK85}
{Hartmann}, L. \& {Kenyon}, S.~J. 1985, \apj, 299, 462

\bibitem[{{Hartmann} \& {Kenyon}(1987)}]{HK87}
---. 1987, \apj, 312, 243

\bibitem[{{Hauschildt} {et~al.}(1999){Hauschildt}, {Allard}, \&
  {Baron}}]{HAB99}
{Hauschildt}, P.~H., {Allard}, F., \& {Baron}, E. 1999, \apj, 512, 377

\bibitem[{{Herbig}(1958)}]{HERBIG58}
{Herbig}, G.~H. 1958, \apj, 128, 259

\bibitem[{{Herbig}(1966)}]{HERBIG66}
---. 1966, Vistas in Astronomy, 8, 109

\bibitem[{{Herbig}(1977)}]{HERBIG77}
---. 1977, \apj, 217, 693

\bibitem[{{Herbig}(2009)}]{HERBIG09}
---. 2009, \aj, 138, 448

\bibitem[{{Herbig} {et~al.}(2003){Herbig}, {Petrov}, \& {Duemmler}}]{HPD03}
{Herbig}, G.~H., {Petrov}, P.~P., \& {Duemmler}, R. 2003, \apj, 595, 384

\bibitem[{{Kenyon} \& {Hartmann}(1991)}]{KH91}
{Kenyon}, S.~J. \& {Hartmann}, L.~W. 1991, \apj, 383, 664

\bibitem[{{Laugalys} \& {Strai{\v z}ys}(2002)}]{LS02}
{Laugalys}, V. \& {Strai{\v z}ys}, V. 2002, Baltic Astronomy, 11, 205

\bibitem[Lodato \& Clarke(2004)]{LC04} 
Lodato, G., \& Clarke, C.~J.\ 2004, \mnras, 353, 841 

\bibitem[{{Malbet} {et~al.}(2005){Malbet}, {Lachaume}, {Berger}, {Colavita},
  {di Folco}, {Eisner}, {Lane}, {Millan-Gabet}, {S{\'e}gransan}, \&
  {Traub}}]{MALBET+05}
{Malbet}, F., {Lachaume}, R., {Berger}, J.-P., {Colavita}, M.~M., {di Folco},
  E., {Eisner}, J.~A., {Lane}, B.~F., {Millan-Gabet}, R., {S{\'e}gransan}, D.,
  \& {Traub}, W.~A. 2005, \aap, 437, 627

\bibitem[{{Millan-Gabet} {et~al.}(2006){Millan-Gabet}, {Monnier}, {Akeson},
  {Hartmann}, {Berger}, {Tannirkulam}, {Melnikov}, {Billmeier}, {Calvet},
  {D'Alessio}, {Hillenbrand}, {Kuchner}, {Traub}, {Tuthill}, {Beichman},
  {Boden}, {Booth}, {Colavita}, {Creech-Eakman}, {Gathright}, {Hrynevych},
  {Koresko}, {Le Mignant}, {Ligon}, {Mennesson}, {Neyman}, {Sargent}, {Shao},
  {Swain}, {Thompson}, {Unwin}, {van Belle}, {Vasisht}, \&
  {Wizinowich}}]{MILLAN-GABET+06b}
{Millan-Gabet}, R., {Monnier}, J.~D., {Akeson}, R.~L., {Hartmann}, L.,
  {Berger}, J.-P., {Tannirkulam}, A., {Melnikov}, S., {Billmeier}, R.,
  {Calvet}, N., {D'Alessio}, P., {Hillenbrand}, L.~A., {Kuchner}, M., {Traub},
  W.~A., {Tuthill}, P.~G., {Beichman}, C., {Boden}, A., {Booth}, A.,
  {Colavita}, M., {Creech-Eakman}, M., {Gathright}, J., {Hrynevych}, M.,
  {Koresko}, C., {Le Mignant}, D., {Ligon}, R., {Mennesson}, B., {Neyman}, C.,
  {Sargent}, A., {Shao}, M., {Swain}, M., {Thompson}, R., {Unwin}, S., {van
  Belle}, G., {Vasisht}, G., \& {Wizinowich}, P. 2006, \apj, 641, 547

\bibitem[{{Miller} {et~al.}(2011){Miller}, {Hillenbrand}, {Covey}, {Poznanski},
  {Silverman}, {Kleiser}, {Rojas-Ayala}, {Muirhead}, {Cenko}, {Bloom},
  {Kasliwal}, {Filippenko}, {Law}, {Ofek}, {Dekany}, {Rahmer}, {Hale}, {Smith},
  {Quimby}, {Nugent}, {Jacobsen}, {Zolkower}, {Velur}, {Walters}, {Henning},
  {Bui}, {McKenna}, {Kulkarni}, {Klein}, {Kandrashoff}, \&
  {Morton}}]{MILLER+11}
{Miller}, A.~A., {Hillenbrand}, L.~A., {Covey}, K.~R., {Poznanski}, D.,
  {Silverman}, J.~M., {Kleiser}, I.~K.~W., {Rojas-Ayala}, B., {Muirhead},
  P.~S., {Cenko}, S.~B., {Bloom}, J.~S., {Kasliwal}, M.~M., {Filippenko},
  A.~V., {Law}, N.~M., {Ofek}, E.~O., {Dekany}, R.~G., {Rahmer}, G., {Hale},
  D., {Smith}, R., {Quimby}, R.~M., {Nugent}, P., {Jacobsen}, J., {Zolkower},
  J., {Velur}, V., {Walters}, R., {Henning}, J., {Bui}, K., {McKenna}, D.,
  {Kulkarni}, S.~R., {Klein}, C.~R., {Kandrashoff}, M., \& {Morton}, A. 2011,
  \apj, 730, 80

\bibitem[{{Pfalzner}(2008)}]{PFALZNER08}
{Pfalzner}, S. 2008, \aap, 492, 735

\bibitem[{{Pott} {et~al.}(2010){Pott}, {Woillez}, {Ragland}, {Wizinowich},
  {Eisner}, {Monnier}, {Akeson}, {Ghez}, {Graham}, {Hillenbrand},
  {Millan-Gabet}, {Appleby}, {Berkey}, {Colavita}, {Cooper}, {Felizardo},
  {Herstein}, {Hrynevych}, {Medeiros}, {Morrison}, {Panteleeva}, {Smith},
  {Summers}, {Tsubota}, {Tyau}, \& {Wetherell}}]{POTT+10}
{Pott}, J., {Woillez}, J., {Ragland}, S., {Wizinowich}, P.~L., {Eisner}, J.~A.,
  {Monnier}, J.~D., {Akeson}, R.~L., {Ghez}, A.~M., {Graham}, J.~R.,
  {Hillenbrand}, L.~A., {Millan-Gabet}, R., {Appleby}, E., {Berkey}, B.,
  {Colavita}, M.~M., {Cooper}, A., {Felizardo}, C., {Herstein}, J.,
  {Hrynevych}, M., {Medeiros}, D., {Morrison}, D., {Panteleeva}, T., {Smith},
  B., {Summers}, K., {Tsubota}, K., {Tyau}, C., \& {Wetherell}, E. 2010, \apj,
  721, 802

\bibitem[{{Quanz} {et~al.}(2006){Quanz}, {Henning}, {Bouwman}, {Ratzka}, \&
  {Leinert}}]{QUANZ+06}
{Quanz}, S.~P., {Henning}, T., {Bouwman}, J., {Ratzka}, T., \& {Leinert}, C.
  2006, \apj, 648, 472

\bibitem[{{Racine}(1968)}]{RACINE68}
{Racine}, R. 1968, \aj, 73, 233

\bibitem[{{Reid} {et~al.}(2009){Reid}, {Menten}, {Zheng}, {Brunthaler},
  {Moscadelli}, {Xu}, {Zhang}, {Sato}, {Honma}, {Hirota}, {Hachisuka}, {Choi},
  {Moellenbrock}, \& {Bartkiewicz}}]{REID+09}
{Reid}, M.~J., {Menten}, K.~M., {Zheng}, X.~W., {Brunthaler}, A., {Moscadelli},
  L., {Xu}, Y., {Zhang}, B., {Sato}, M., {Honma}, M., {Hirota}, T.,
  {Hachisuka}, K., {Choi}, Y.~K., {Moellenbrock}, G.~A., \& {Bartkiewicz}, A.
  2009, \apj, 700, 137

\bibitem[{{Reipurth} \& {Aspin}(2004)}]{RA04}
{Reipurth}, B. \& {Aspin}, C. 2004, \apjl, 608, L65

\bibitem[{{Reipurth} \& {Aspin}(2010)}]{RA10}
{Reipurth}, B. \& {Aspin}, C. 2010, in Evolution of Cosmic Objects through
  their Physical Activity, ed. {H.~A.~Harutyunian, A.~M.~Mickaelian, \&
  Y.~Terzian}, 19--38

\bibitem[{{Vorobyov} \& {Basu}(2005)}]{VB05}
{Vorobyov}, E.~I. \& {Basu}, S. 2005, \apjl, 633, L137

\bibitem[{{Zhu} {et~al.}(2007){Zhu}, {Hartmann}, {Calvet}, {Hernandez},
  {Muzerolle}, \& {Tannirkulam}}]{ZHU+07}
{Zhu}, Z., {Hartmann}, L., {Calvet}, N., {Hernandez}, J., {Muzerolle}, J., \&
  {Tannirkulam}, A. 2007, \apj, 669, 483

\bibitem[{{Zhu} {et~al.}(2008){Zhu}, {Hartmann}, {Calvet}, {Hernandez},
  {Tannirkulam}, \& {D'Alessio}}]{ZHU+08}
{Zhu}, Z., {Hartmann}, L., {Calvet}, N., {Hernandez}, J., {Tannirkulam}, A., \&
  {D'Alessio}, P. 2008, \apj, 684, 1281

\bibitem[{{Zhu} {et~al.}(2009){Zhu}, {Hartmann}, {Gammie}, \&
  {McKinney}}]{ZHU+09}
{Zhu}, Z., {Hartmann}, L., {Gammie}, C., \& {McKinney}, J.~C. 2009, \apj, 701,
  620

\end{thebibliography}
\end{document}